\documentclass[graybox]{svmult}

\usepackage{type1cm}        
\usepackage{makeidx}         
\usepackage{graphicx}        
\usepackage{multicol}        
\usepackage[bottom]{footmisc}
\usepackage{todonotes}
\usepackage{rotating}
\usepackage{hyperref}

\usepackage{newtxtext}       %
\usepackage{newtxmath}       
\usepackage{verbatim}
\usepackage{lscape}
\usepackage{afterpage}

\makeindex             

\begin{document}

\title*{Designing a Syllabus for a Course on Empirical Software Engineering }
\author{Paris Avgeriou, \\ 
Nauman bin Ali, \\ 
Marcos Kalinowski,\\
Daniel Mendez}
\institute{Paris Avgeriou \at University of Groningen \email{p.avgeriou@rug.nl} \and  Nauman bin Ali \at Blekinge Institute of Technology \email{nauman.ali@bth.se}\and Daniel Mendez \at Blekinge Institute of Technology and
fortiss \email{daniel.mendez@bth.se} \and Marcos Kalinowski \at Pontifical Catholic University of Rio de Janeiro, \email{kalinowski@inf.puc-rio.br}}

\maketitle

\abstract{Increasingly, courses on Empirical Software Engineering research methods are being offered in higher education institutes across the world, mostly at the M.Sc. and Ph.D. levels. While the need for such courses is evident and in line with modern software engineering curricula, educators designing and implementing such courses have so far been reinventing the wheel; every course is designed from scratch with little to no reuse of ideas or content across the community. Due to the nature of the topic, it is rather difficult to get it right the first time when defining the learning objectives, selecting the material, compiling a reader, and, more importantly, designing relevant and appropriate practical work. This leads to substantial effort (through numerous iterations) and poses risks to the course quality.  \\
This chapter attempts to support educators in the first and most crucial step in their course design: creating the syllabus. It does so by consolidating the collective experience of the authors as well as of members of the Empirical Software Engineering community; the latter was mined through two working sessions and an online survey. Specifically, it offers a list of the fundamental building blocks for a syllabus, namely course aims, course topics, and practical assignments. The course topics are also linked to the subsequent chapters of this book, so that readers can dig deeper into those chapters and get support on teaching specific research methods or cross-cutting topics. Finally, we guide educators on how to take these building blocks as a starting point and consider a number of relevant aspects to design a syllabus to meet the needs of their own program, students, and curriculum. 
}

\section{Introduction}
\label{sec:intro}

In the previous chapter (Editorial Introduction), we established the need for offering high-quality courses on Empirical Software Engineering (ESE) to educate students in higher education institutions (mostly at MSc and PhD levels) or train practitioners working in industrial Research and Development. We have also elaborated on several challenges that educators face when designing such courses. These challenges are partly because the topic is still rather young, and we do not yet have dedicated textbooks or reusable syllabi to guide course designers. However, the challenges are also due to the complex nature of this topic offering ample freedom for course designers to tailor their course. This freedom is both a blessing and a curse: a blessing as it allows one to cater to the needs of individual programs, curricula, and target audiences; a curse, because such tailoring takes many iterations before an educator gets it right.

To understand how such courses are designed and run by educators, we organized two consecutive working sessions during the annual meetings of the International Software Engineering Research Network (ISERN) in 2021 and 2022. The second session was combined with an online questionnaire. The two working sessions were attended by 31 unique participants while the questionnaire was filled in by 27 respondents (15 of them overlapping with the working session participants). All aforementioned participants and respondents were affiliated (at the time) with academic institutions from Europe, and North and South America.

During these working sessions and the online questionnaire, we collected a set of: a) Course Aims, b) Course Topics, and c) Course Assignments. We also had the chance to discuss with the participants about their experiences on course design. One common but striking remark from those sessions was that, by and large, every educator "reinvents the wheel". There is little to no reuse of any aspect of course design in ESE. This not only leads to substantial effort every time a course is (re-)designed, but it also entails significant variation in quality.

To address this problem, this chapter aims at offering some building blocks that can provide a starting point for designing a course in ESE\footnote{While this chapter refers to education in Software Engineering, educators can draw inspiration for similar Research Methods courses in related disciplines like Information Systems.}. To be more precise, this chapter offers a collection of course aims, topics and assignments; these are the three fundamental elements of a syllabus. Furthermore, the chapter includes guidelines on how to create a syllabus as well as a number of other pertinent aspects (e.g. ethics approval, datasets, tools, and prerequisites). The intent is that course designers, by following these guidelines, can select elements from the provided building blocks and refine them to create a syllabus to match their curriculum and program needs, student population, and, of course, personal teaching style and philosophy. By mixing and matching the elements of this chapter, we hope that course designers can reuse the collective wisdom of the ESE community, as well as their long experience in designing and teaching courses. That should effectively lead to fewer iterations in course design and better quality assurance in new or re-designed courses.

We have intentionally not provided a single, one-size-fits-all syllabus for a course in ESE, as we believe each such course is unique and requires a tailored syllabus. We thus advise educators to perform a step-wise refinement and customization of the offered building blocks to address the context of their course and of course their background. To further assist them in this task, we: a) offer a simple set of guidelines and discuss pertinent aspects to consider in Section \ref{sec:misc}; b) link the Course Topics to other chapters in the book as well as external sources in Section \ref{sec:topics}. These links can allow course designers to delve deeper into the specific topics they want to include in their course, as those are covered in later chapters or elsewhere. For example, if a course designer has selected to teach experimentation and give an assignment to design and run an experiment, the chapter of Vegas and Juristo, entitled ``A Course on Experimentation in Software Engineering: Focusing on Doing'', provides excellent details, tips, and examples.

The rest of this chapter is structured as follows: Section~\ref{sec:LO} lists a wide range of \textit{Course Aims} that are pertinent to ESE courses; Section~\ref{sec:topics} presents an inclusive list of \textit{Topics} that can be taught in such a course, linking them to the rest of the chapters of this book, as well as other sources; Section~\ref{sec:assignments} discusses potential \textit{Assignments} to help students understand the theory covered in the topics. 
Finally, Section~\ref{sec:misc} proposes guidelines on how to use the aforementioned elements (course aims, topics, assignments) to create an actual syllabus, and it touches upon various miscellaneous aspects such as ethics approval or tools and datasets to be considered for the syllabus.

\section{Course Aims}
\label{sec:LO}

The following list of Course Aims is meant to cover a wide range of goals for an ESE course at the undergraduate or postgraduate level. As mentioned, this list is based on data collected from experts during two ISERN working sessions and from an online questionnaire. It is not meant to be an exhaustive list and it does not distinguish between undergraduate and postgraduate levels; we leave it at the discretion of the reader to reuse, refine, or add course aims according to the profile of the education program (\textit{e.g.}, technical university vs. classical university, faculty of natural sciences vs. faculty of computer science, program in Computer Science vs. program in Software Engineering), the level of the students, as well as the fit within the aims of the education program. These course aims can subsequently be refined into Learning Objectives by making them more specific and measurable.

After completing the course on ESE, students must be able to:
\begin{itemize}

\item\textbf{Understand philosophical stances}: list the different schools of thought in ESE research (such as positivism, interpretivism, critical theory, pragmatism, or realism), how they complement each other, and position research papers in this landscape.

\item\textbf{Understand the importance and role of empirical research}: explain the contribution of empirical research alongside other methods (such as formal methods used for rationalist arguments) in advancing the state of the art in software engineering and recognize the importance of evidence for software practitioners.

\item\textbf{Understand the basic methods and components of empirical research}: explain the contribution of primary and secondary studies for evidence-based software engineering, define the nature and aim of typical research methods (\textit{e.g.}, surveys, case studies, experiments, or systematic literature reviews), and explain their core parts (such as research design, threats to validity, data collection, and analysis) as well as the difference between quantitative and qualitative data.

\item\textbf{Understand how to read, interpret and summarise research papers}: using empirical standards and best practices for peer review, describe the typical structure of a research paper, identify the main parts (e.g., study design, data collection, and analysis), interpret both quantitative and qualitative results and describe them to both specialists and laymen.

\item\textbf{Classify and explain the problem or question under investigation}: motivate a gap in the state of research and practice, and formulate research questions of different categories (e.g. exploratory, base-rate, causality), that are precise enough and can be answered through evidence.

\item\textbf{Select a research design}: use the formulated research questions to select a research method (or a mix of methods) that can collect the evidence to answer them.

\item\textbf{Design and execute an empirical study}: complete the study design according to specific research methods (including, for instance, Quantitative and Qualitative data), and conduct the study following that design.

\item\textbf{Perform data collection}: execute the data collection part of the study design by developing data collection instruments or pertinent tools for several data sources to collect both quantitative and qualitative data.

\item\textbf{Perform data analysis}: analyse the collected data (both quantitative and qualitative) by visualizing (if possible, otherwise suitably representing), exploring, and subsequently reflecting upon, interpreting, and explaining the data.

\item\textbf{Communicate research results}: report the results of an empirical study by following established guidelines both in written form and orally.

\item\textbf{Assemble a replication package}: systematically collect all datasets, scripts, and other artifacts into a single package (depending also on the type of study), describe how to use it, and publish it in a permanently and publicly available online repository.

\item\textbf{Develop critical thinking and a scientific attitude}: follow a systematic process to conceptualise, analyse, synthesise, and evaluate collected data.

\item\textbf{Develop theories}: use the evidence collected from the empirical study to develop theories that advance the state of the art in software engineering.

\item\textbf{Understand practical aspects of research}: derive implications from research results for both researchers and practitioners and devise a plan to transfer knowledge to industry.

\item\textbf{Understand ethical aspects of research}: analyse the potential ethical implications of the study (e.g. treatment of human subjects, confidentiality of data) and devise a plan to address any ethical concerns, potentially to be approved by an ethics board.

\end{itemize}
\section{Topics}
\label{sec:topics}

The following is a list of topics, based on the aforementioned input from the experts and clustered into five broad themes. The list is not meant to be exhaustive, but indicative of the most \textit{commonly} used topics in ESE courses.

\begin{enumerate}

\item Fundamentals of ESE - the general topics in this category are the theoretical underpinnings of ESE and help set the foundations for the research methods to be covered in a course on ESE.  We have relied on the terminology proposed by Wohlin and Ayb{\"{u}}ke \cite{WohlinA15} for the topics in this category.
\begin{itemize}
    \item Research approach/Philosophical stances - A philosophical stance refers to a researcher's underlying philosophical perspective about research (or a combination of perspectives). It indicates a researcher's epistemological and ontological assumptions and impacts how researchers approach an investigation, interpret the findings, and draw conclusions. Some common philosophical stances in software engineering research \cite{EasterbrookSSD08} include: (1) Post-positivism, (2) Interpretivism/Constructivism, (3) Critical Theory, and (4) Pragmatism.
    \item Research logic (inductive, deductive or abductive research) - As discussed in Chapter 1, empirical research follows an iterative process between theory building and empirical approaches through induction, deduction or abduction (please see also ~\cite{mendez19EMSEInterdiscipline}). The chapter by Klaas Jan Stol in this book, titled ``Teaching Theorizing in Software Engineering Research'' discusses how findings can feed back to the theories and the overall Software Engineering body of knowledge.
    \item Research outcome (basic or applied research) - Research outcome can be new knowledge contributions or improved solutions for practice. The chapter by Rodrigo Falcão et al. titled ``Experiences in Using the V-Model as a Framework for Applied Doctoral Research'' presents a framework that discusses the role and interplay of the research outcomes of basic and applied research.
    \item Research design - the choices made in creating a research design including research methodology, data collection and analysis methods. These decisions and alternatives are discussed in the Chapter by Jefferson Molleri and Kai Petersen titled ``Teaching Research Design in Software Engineering''. 
    \item Research process -- Broadly, there are three types of research processes, namely quantitative, qualitative, and mixed research \cite{WohlinA15}. The book covers several methods that are used in these research processes. Chapters that cover quantitative research range from experimentation by Juristo and Vegas (entitled ``A Course on Experimentation in Software Engineering: Focusing on Doing'') to simulation by Franca et al. (entitled ``Teaching Simulation as a Research Method in Empirical Software Engineering''). Chapters on qualitative data analysis by Treude (entitled ``Qualitative Data Analysis in Software Engineering: Techniques and Teaching Insights'') and on ethnographies by Dittrich et al. (entitled ``Teaching and Learning Ethnography for Software Engineering Contexts'') cover aspects relevant to qualitative research.

\end{itemize}

\item Typical research methods - The following topics cover a variety of research methods commonly employed in software engineering research. It is wise to include a range of methods in a course; although it may not be feasible to include all of them, we advise covering both quantitative and qualitative methods.
    \begin{itemize}
        \item Action research - a strategy for investigating a problem or question within a specific real-world setting to bring about a change (see the chapter of Miroslaw Staron, titled ``Teaching Action Research'', as well as the chapter by Yvonne Dittrich et al. titled ``Action Research with Industrial Software Engineering – An Educational Perspective'').
        \item Case study research - a strategy for an in-depth investigation to develop a better understanding of a phenomenon of interest in a specific real-world setting (see the chapter by Stefan Wagner titled ``Teaching Case Study Research'').
        \item Design science - a strategy for software engineering research that emphasizes practical application through the creation and evaluation of artifacts (e.g., templates, systems, and processes) in a real-world setting (see the chapter by Oscar Pastor et al. titled ``Teaching Design Science as a Method for Effective Research Development'').
        \item Ethnography -- an empirical investigation strategy to study and understand the people and communities and cultures in their natural settings (see the Chapter by Yvonne Dittrich et al. titled ``Teaching and Learning Ethnography for Software Engineering Contexts'').
        \item Experimentation -- an investigation strategy to test hypothesis regarding causal relations in a controlled setting (see the chapter by Natalia Juristo and Sira Vegas titled ``A Course on Experimentation in Software Engineering: Focusing on Doing'').
        \item Simulation -- an investigation strategy that relies on imitating the behavior of a system in a mathematical model and then experimenting with the model (see the Chapter by Breno Bernard Nicolau de França et al. titled "Teaching Simulation as a Research Method in Empirical Software Engineering").
        \item Survey research -- used to investigate the opinions, attitudes, and characteristics of a population by collecting data from a large sample of the population (see the chapter by Marcos Kalinowski et al. titled ``Teaching Survey Research in Software Engineering'').
        \item Repository mining -- a form of archival analysis where the existing data and artifacts are analyzed for insights, patterns, and knowledge about software development (see the chapter by Zadia Codabux et al. titled ``Teaching Mining Software Repositories'').
    \end{itemize}
    \item Operationalization methods - these topics deal with executing empirical methods through data collection, analysis and measurement.
    \begin{itemize}
        \item Data collection methods (like questionnaires, interviews, observations, archival research, simulation, focus groups \cite{WohlinA15}) - Several of these methods are discussed briefly in various chapters of the book, e.g., for simulation see Franca et al.'s chapter on simulation, for observations see Dittrich et al.'s chapter on Action Research, and for questionnaires see Kalinowski et al.'s chapter on Surveys. 
    \item Measurement - this is a fundamental topic in ESE research, encompassing the measurement and prediction of the attributes of the software processes, products, and practitioners. The Chapter by Paul Ralph et al. titled ``Teaching Software Metrology: The Science of Measurement for Software Engineering'' elaborates in-depth on the nuances of teaching this topic.
        \item Data analysis 
        \begin{itemize}
            \item Quantitative data analysis - covers descriptive and inferential statistics (\textit{e.g.}, hypothesis testing, Bayesian analysis, confidence interval estimation, machine learning models). Several excellent guidelines for quantitative data analysis exist, including guidelines in the context of controlled experiments \cite{wohlin2012experimentation}, when comparing algorithms \cite{arcuri2011practical}, and on Bayesian statistics \cite{Furia2021}.
        
            \item Qualitative data analysis - covers the sense-making and analysis of qualitative data. The Chapter by Christoph Treude titled ``Qualitative Data Analysis in Software Engineering: Techniques and Teaching Insights'' provides a thorough coverage of teaching this aspect.
    \end{itemize}
    \end{itemize}
\item Cross-cutting concerns - such aspects are not specific to a particular research method, but they need to be addressed in any type of empirical study.
\begin{itemize}
    \item Ethics - covers, among others, research ethics, authorship, consent, and conflicting interests. In this book, where applicable, each chapter briefly discusses the relevant ethical concerns for the specific research method it covers. For a more general discussion of ethics, please see \cite{Gotterbarn14,AydemirD18}; interested readers can also consult \cite{Towell03} for teaching ethics.
    \item Sustainability, -- covers the principles and practices adhering to which can help improve sustainability\footnote{Sustainability in this context is viewed broadly, for example aligning with the UN Sustainable Development Goals \url{https://sdgs.un.org/goals}} in research. The contribution by Birgit Penzenstadler, titled ``Sustainability Competencies Informing Research Strategies for Software Engineering: A Personal Experience Report'' gives a personal reflection on teaching this topic.
\end{itemize}

\item Scientific literature - this category covers the essence of what constitutes scientific literature, the peer review process, and both systematic and non-systematic literature reviews.
\begin{itemize}
        \item Reading and reviewing research papers - not directly covered in the book but interested readers can consider consulting the ACM empirical standards for software engineering \cite{ralph2020empirical}.
        \item Secondary and tertiary studies - we have two related chapters here: (1) a chapter by Sebastian Baltes and Paul Ralph titled ``Teaching Literature Reviews in Software Engineering Research'' and (2) a chapter by Marcela Genero and Mario Piattini, titled ``Teaching Systematic Literature Reviews: Strategies and Best Practices''.
    \end{itemize}
\end{enumerate}

The Course Aims that an educator selects for a course in ESE should be sufficiently covered by the corresponding Topics. One of the possible mappings between them is given in Table \ref{tab:mapCourseAims2Topics}; this mapping can be used when checking the coverage of Course Aims by Topics. We also use it within the proposed guidelines to design a syllabus in Section \ref{sec:misc}. 

\begin{landscape}
\begin{table}
\centering
\caption{Mapping between Course Aims and Topics}
\label{tab:mapCourseAims2Topics}
\begin{tabular}{|p{3cm}|p{1.3cm}|p{1.3cm}|p{1.3cm}|p{0.8cm}|p{1.8cm}|p{1.7cm}|p{1.3cm}|p{1.3cm}|p{1.3cm}|p{1.3cm}|p{1.3cm}|}
\hline
Course Aims / Topics & Research approach  & Research logic & Research outcome & Ethics & Sustainability & Measurement & Research design & Research process & Typical empirical strategies & Data analysis  & Scientific literature \\
\hline
Understand philosophical stances & x &  &  & x &  &  &  &  &  &  &  \\
\hline
Understand the importance and role of empirical research & x & x & x & x &  &  &  &  &  &  &  \\
\hline
Understand the basic methods and components of empirical research &  &  &  &  &  &  & x & x & x &  &  \\
\hline
Classify and explain the problem under investigation & x & x &  & x &  &  &  &  &  &  &  \\
\hline
Select a research design &  &  &  &  &  & x & x &  &  &  &  \\
\hline
Design and execute an empirical study &  &  &  & x & x &  &  & x & x &  &  \\
\hline
Communicate research results &  &  &  & x &  &  &  &  &  &  & x \\
\hline
Perform data collection &  &  &  & x &  &  &  &  &  &  &  \\
\hline
Perform data analysis &  &  &  &  &  & x &  &  &  & x &  \\
\hline
Assemble a replication package &  &  &  &  &  &  &  & x &  &  &  \\
\hline
Develop critical thinking and a scientific attitude & x & x &  &  &  &  &  &  &  &  &  \\
\hline
Develop theories &  &  & x &  &  &  &  &  &  &  &  \\
\hline
Understand practical aspects of research &  &  & x &  &  &  &  &  &  &  & x \\
\hline
Understand ethical aspects of research &  &  &  & x &  &  & x & x &  & x &  \\
\hline
Understand how to read, interpret and summarise research papers & x & x &  &  &  &  &  &  &  &  & x \\
\hline

\end{tabular}

\end{table}
\end{landscape}

\section{Course Assignments}
\label{sec:assignments}

Practical assignments play an important role in teaching ESE, helping to bridge the theoretical knowledge of the empirical methods with their application in practice. Assignments should provide students with hands-on experience in the various facets of empirical research, from designing experiments and conducting systematic literature reviews to building instruments for data collection or performing data analysis. By engaging in these practical activities, students not only reinforce their learning but also develop essential skills such as critical thinking, problem-solving, and effective communication. Like teaching Software Engineering, a course on ESE is fundamentally based on `learning by doing'.

The following list provides examples of practical assignments based on the data collected from the aforementioned working sessions and questionnaire, \textit{in alphabetical order}. The chapters elaborating on the different empirical strategies and their supplemental materials provide more specific examples and will be helpful for educators when designing practical assignments for their courses.

\begin{itemize}

\item Adapt an Experiment - students modify an existing experiment to either suit a different context or test a new hypothesis. This encourages an understanding of experimental design, variables, and replication. It also helps students to grasp the adaptability of experimental methods and the importance of contextual factors in empirical research.
\item Assign Group Activities - multiple groups apply different empirical methods ({e.g.}, different systematic literature review search strategies) to investigate the same topic and compare and discuss the outcomes, fostering a comparative understanding of the methods.
\item Conduct a Secondary Study - students perform a secondary study (\textit{e.g.}, systematic review of literature, systematic mapping study) to identify, analyse, and synthesise relevant research on a specific topic. This assignment enhances skills in critical review and research synthesis.
\item Analyse Data  - students analyse a given dataset using statistical or qualitative methods to draw meaningful inferences. This kind of assignment teaches students how to apply statistical analysis techniques and qualitative methods to real-world data, emphasizing the interpretation and communication of results. A variant of this type of assignment is to perform a replication of an existing study by taking its dataset and reproducing the analysis; this allows to confirm (or not) the findings of that study and discuss/explain any discrepancies.
\item Collect and Analyse Data - students collect data and analyse it using predetermined methods, emphasising also the linkage between data gathering methods and the quality of research findings.
\item Design and Execute an Empirical Study - students design and carry out an empirical study following a method of their choice. For example, in the case of an experiment, students can gain hands-on experience from hypothesis formulation to data collection and analysis, strengthening their understanding of experimental designs. Alternatively, students can design and execute scoped-down versions of this assignment type by executing for example pilot studies, or only performing study design. It is advisable to make available a replication package with the study design, as well as any data, scripts, tools, instruments etc.
\item Develop a Systematic Literature Review Protocol - students specify a reproducible protocol for systematic reviews to sharpen their secondary research planning skills.
\item Participate in Reading Assignment - students participate in reading assignments and group discussions of exemplary papers to develop their analytical skills and deepen their understanding of high-quality research.
\item Present Research - students enhance their presentation and critical thinking skills by presenting scientific topics to the class, such as summaries of research papers or critical analyses of a set of papers.
\item Mine Repositories - students analyse software repositories to uncover patterns and insights, getting familiar with the methodologies and tools necessary for mining extensive data sets.
\item Perform a Toy Case Study - students design, conduct, and analyze a case study centered around a toy problem (a scaled-down, simplified enactment of a software engineering problem).
\item Write a Research Proposal - students outline a research proposal where they must define clear research questions, propose a robust empirical research methodology, and discuss the potential impact and contributions of their proposed research to the field of software engineering.
\end{itemize}

Clearly, some of these assignments are larger in scope and effort than others. For example, the assignment ``Design and Execute an Empirical Study'' encompasses many of the other types of assignments. As another example, the assignment ``Conduct a Secondary Study'' is rather effort-intensive and can easily last an entire semester (if not more). Similarly, some of these assignments are better suited for individual students, while others can be optimally done in groups of students; this also depends on the level of the students (BSc, MSc, PhD, or trainee). We advise educators to take this list as an inspiration and customize each type of assignment by emphasizing particularly the timing constraints and the effort students are expected to spend on each course component (lectures, tutorials, self-studying, practical work). 

Moreover, we would like to emphasize that the required effort for any kind of assignment should be realistic and match the difficulty/complexity of the assignment and the level of the student. Educators should not give assignments that are so difficult and effort-intensive, that students end up cutting corners to be able to pass the course. Instead, educators should uphold high standards of quality and rigor so that students understand the importance of quality in both the research process and the outcome.

Finally, educators should look at the practical assignments as a means of achieving the Course Aims. If there is a mismatch between Course Aims and Assignments, then one of them needs refinement. One of the possible mappings between them is given in Table \ref{tab:MapCourseAims2Assignments}; this mapping can be used when checking the achievement of Course Aims through Assignments. We also use it within our guidelines to design a syllabus in Section \ref{sec:misc}. To give a more concrete feeling of what actual assignments look like, we have appended some assignments from actual courses in the online resources for this chapter (see \url{https://zenodo.org/doi/10.5281/zenodo.11544897}). 


\begin{landscape}
\begin{table}
\centering
\caption{Mapping between Course Aims and Assignments}
\label{tab:MapCourseAims2Assignments}
\begin{tabular}{|p{3cm}|p{1.1cm}|p{1.3cm}|p{1.3cm}|p{1.1cm}|p{1.4cm}|p{1.4cm}|p{1.3cm}|p{1.3cm}|p{1.2cm}|p{1.3cm}|p{1cm}|p{1.2cm}|}
\hline
Course Aims / Assignments & Adapt an Experiment & Assign Group Activities & Conduct a Secondary Study & Conduct Data Analysi & Conduct Data Collection and Analysis & Design and Execute an Empirical Study & Develop a Systematic Literature Review Protocol & Participate in Reading Assignment & Present Research & Mine Repositories & Perform a Toy Case Study & Write a Research Proposal \\
\hline
Understand philosophical stances & x &  &  &  &  & x &  &  &  &  &  & x \\
\hline
Understand the importance and role of empirical research &  &  & x &  &  &  &  & x &  &  &  &  \\
\hline
Understand the basic methods and components of empirical research &  &  & x &  &  &  &  &  &  &  &  & x \\
\hline
Classify and explain the problem under investigation &  &  & x &  &  &  &  &  &  &  &  &  \\
\hline
Select a research design &  &  &  &  &  & x &  &  &  &  &  &  \\
\hline
Design and execute an empirical study & x & x &  &  &  & x &  &  &  &  & x &  \\
\hline
Communicate research results &  &  &  &  & x &  &  &  & x &  &  &  \\
\hline
Perform data collection &  &  & x &  & x & x &  &  &  & x &  &  \\
\hline
Perform data analysis &  &  & x & x &  & x &  &  &  & x &  &  \\
\hline
Assemble a replication package &  &  & x &  &  &  &  &  &  &  &  &  \\
\hline
Develop critical thinking and a scientific attitude & x & x & x &  &  &  & x & x &  &  &  &  \\
\hline
Develop theories &  &  &  &  &  &  &  &  &  &  &  &  \\
\hline
Understand practical aspects of research & x &  &  &  &  &  &  &  &  &  &  & x \\
\hline
Understand ethical aspects of research &  &  &  &  & x & x &  &  &  &  &  & x \\
\hline
Understand how to read, interpret and summarise research papers &  &  & x &  &  &  & x & x &  &  &  & x \\
\hline 
\end{tabular}

\end{table}
\end{landscape}


\section{Putting it all together to create a syllabus}
\label{sec:misc}

The Course Aims, Topics and Assignments described in the previous sections are a starting point, but how does one go about actually creating a syllabus for one's course? We recommend the following steps:
\begin{enumerate}
    \item Select the Course Aims that fit the course, the overall curriculum and the student population. Subsequently, derive concrete Learning Objectives that are SMART (Specific, Measurable, Achievable, Relevant, and Time-bound). For example, the course aim ``Analyse data'' can be refined into the learning objective ``Apply the Constant Comparison method to a collection of interview transcripts''. We strongly advise educators against simply adopting a subset of the course aims, but to really reflect on how to tailor them to their own context and make them specific and measurable. 
    \item Based on the formulated Learning Objectives, select the Topics to be covered during the course; the mapping between Course Aims (from which the Learning objectives are derived) and Topics in Table \ref{tab:mapCourseAims2Topics} is a good starting point. Follow the references to the chapters of the rest of the book or to external sources to select teaching material, slides, and reading lists.
    \item Based on the formulated Learning Objectives and the selected Topics, select Assignments that optimally fit the educator's context, also choosing appropriate datasets, tools, and assessment rubrics; the mapping between Course Aims (from which the Learning objectives are derived) and Assignments in Table \ref{tab:MapCourseAims2Assignments} is a good starting point. It is important that the syllabus explains to the students how the Assignments achieve the Learning Objectives, so that they understand the 'why' and not only the 'how' (e.g., why they can develop a theory from a specific dataset and why not). 
\end{enumerate}

The aforementioned steps touch upon a number of decisions that an educator needs to make to complete their course syllabus. These decisions include but are not limited to, the following:\\
\textbf{Ethics approval}. The approval of the ethics board varies widely. Some universities have it as mandatory, while in other cases, it acts as an obstacle as it takes quite long so it's difficult to fit it into the course timeline. 
We advise that, as a minimum, there is a discussion on research ethics and that students are made well aware of the ethical issues in their assignments and report on them. If the formal approval by the university ethics board is infeasible in the course time frame, an ad-hoc ethics approval committee can be formed in the class either with the students themselves, and/or with the educator and Teaching Assistants.  \\
\textbf{Reading list}. Each topic covered in the course needs to be accompanied by examples of good papers (for instance, by using existing SLRs and relying on their quality assessment score to identify high quality papers for a given topic)
, as well as examples of papers to be critically reviewed to identify points for improvement. Students should be encouraged to reflect and understand the strengths of the papers they read, but also to have a critical attitude towards identifying elements of an empirical study that can be optimized (e.g. threats to validity). The latter should always be performed professionally and constructively. 
Finally, it is advisable to link the assignments with the reading list, so that students have a concrete incentive to read about a topic in depth (e.g., read an authoritative source on case study research, before performing a case study). \\
\textbf{Datasets}. It is a good practice for educators to provide examples of datasets from the replication packages of specific papers or collections from open science repositories (e.g. https://zenodo.org/communities/seacraft/). Of course, educators may also have their own datasets that students would need to analyze as part of their assignments. In that case, it is very important for educators to provide clear guidance on how students can analyse the dataset. \\
\textbf{Open Science}. Make principles and practices of open data, open material, and open source explicit, including available and recommended license models. In combination with the datasets as discussed before, this also helps the students understand the features of a good replication package that follows sound Open Science principles. It is important to make it clear to students that different types of studies lead to different replication packages; for example, a quantitative study on Open Source data can generate very large and public datasets, while an ethnographic or action research study may require data redaction and/or making only parts of data available.\\
\textbf{Tools}. There are many tools available for conducting the practical assignments, varying based on the topics covered within the course. These include generic tools such as reference management tools for systematic reviews, qualitative data analysis software tools, and tools for surveys. Some more specialized tools are also available (\textit{e.g.}, to support the automation of systematic reviews~\cite{felizardo2020automating}). We recommend focusing on the underlying concepts and making the course as tool-agnostic as possible.\\
\textbf{Prerequisites}. Make prerequisites explicit for the course, e.g., mathematics and statistics required for the students to be able to work on the assignments. It is equally important to make sure the course follows up on the background of the students. For example, the course may require that the students have a background in Software Engineering, e.g. by having followed courses from a Software Engineering curriculum and having participated in a capstone project. This of course depends also on the course level: BSc, MSc, PhD, life-long learning, corporate training etc. Finally, the expectations on previous research experience should be explicit, e.g. whether the students have already completed a research project such as a research internship or a BSc thesis.\\
\textbf{Teaching Methods}. A course in ESE can utilise different teaching methods, including but not limited to lectures, tutorials, seminars, practical work, group or individual assignments, and lab work. While we cannot prescribe a one-size-fits-all configuration of these teaching methods, we strongly advise that the course has a very practical, hands-on nature, where knowledge and skills are acquired mostly by doing rather than by listening to lectures or studying the reading list. For example, having tutorials on how to apply the research methods in practice and labs on how to perform data collection and analysis, has worked very well, in the experience of the authors. Coaching students for their assignments on an individual or group basis and with a high frequency (e.g. weekly) also ensures that students have several intermediate deadlines and get feedback early and often.\\
\textbf{Grading Rubric}. Depending on the level of the course (e.g., graduate vs. undergraduate), some universities have a grading system while others simply require attendance. In the former case, a syllabus needs to provide a grading rubric with an explicit and detailed break-down of how the grade is composed. Typically this rubric takes into account several components such as Quality of deliverables (e.g. study design and study results, as well as Open Science artifacts), Execution (e.g. choice of research method and formulating research questions, background and related work, data collection, data analysis, results, threats to validity), Process (e.g., teamwork and meeting deadlines, motivation, initiative and creativity, following feedback) and Presentations (e.g., content and style of presentations, answering questions). \\
\textbf{Course structure}. A week-by-week breakdown of all the teaching methods should be elaborated on the syllabus, together with all important dates and deadlines. A semester-long course has a very different structure than an intensive 5-day workshop and supports very different Learning Objectives. Educators need to ensure their set of Learning Objectives can be met within the timeframe and budgeted effort.

In the online resources for this chapter, found under \url{https://zenodo.org/doi/10.5281/zenodo.11544897}, we also provide a number of Course Syllabi to be used as examples. 

\bibliographystyle{spmpsci}
\bibliography{literature}
\end{document}